# Transparent Privacy is Principled Privacy

**Ruobin Gong**[1]

[1]Department of Statistics, Rutgers University, New Brunswick, New Jersey, United States of America

**ABSTRACT**

In a technical treatment, this article establishes the necessity of *transparent* privacy for drawing unbiased statistical inference for a wide range of scientific questions. Transparency is a distinct feature enjoyed by *differential privacy*: the probabilistic mechanism with which the data are privatized can be made public without sabotaging the privacy guarantee. Uncertainty due to transparent privacy may be conceived as a dynamic and controllable component from the total survey error perspective. As the 2020 U.S. Decennial Census adopts differential privacy, constraints imposed on the privatized data products through optimization constitute a threat to transparency and result in limited statistical usability. Transparent privacy presents a viable path toward *principled* inference from privatized data releases, and shows great promise toward improved reproducibility, accountability, and public trust in modern data curation.

---

## Media Summary

When conducting statistical analysis using privacy-protected data, the transparency of the privacy mechanism is a crucial ingredient for trustworthy inferential conclusions. The knowledge about the privacy mechanism enables accurate uncertainty quantification and ensures high statistical usability of the data product. This article discusses the key statistical considerations behind transparent privacy, which leads to improved reproducibility, accountability, and public trust. It weighs a few challenges to transparency that emerge from the adoption of differential privacy by the 2020 U.S. Decennial Census.

---

## 1. Introduction

The Decennial Census of the United States is a comprehensive tabulation of its residents. For over two centuries, the census data supplied benchmark information about the states and the country, helped guide policy decisions, and provided crucial data in many branches of the demographic, social, and political sciences. The census aims to truthfully and accurately document the presence of every individual in the United States. The fine granularity of the database, compounded by its massive volume, portrays American life in great detail.

The U. S. Census Bureau is bound by Title 13 of the United States Code to protect the privacy of individuals and businesses who participate in its surveys. These surveys contain centralized and high-quality information about the respondents. If disseminated without care, they might pose a threat to the respondents' privacy. The Bureau implements protective measures to reduce the risk of inadvertently disclosing confidential information. The first publicly available documentation of these methods dates back to 1970 (McKenna, 2018). Until the 2010 Census, statistical disclosure limitation (SDL) mechanisms deployed by the Census Bureau relied to a large extent on table suppression and data swapping, occasionally supplemented by imputation and partially synthetic data. These techniques restricted the verbatim release of confidential information through the data products. However, they do not offer an exposition of privacy protection as a goal in itself. What does the SDL mechanism aim to achieve, and how do we know whether it is actually working? The answers to these questions are not definitive. In particular, the extent of an SDL mechanism's intrusiveness on data usability is not measured and weighed against the extent of privacy protection it affords. We now understand that many traditional SDL techniques are not just ambiguous in definition, but defective in effect, for they can be invalidated by carefully designed attacks that leverage modern computational advancements and auxiliary sources of open access information (see, e.g., Dinur & Nissim, 2003; Sweeney, 2002). With the aid of publicly available data, the Census Bureau attempted a 'reidentification' attack on its own published 2010 Census tabulations, and was successful in faithfully reconstructing as much as 17% of the U.S. population, or 52 million people at the level of individuals (Abowd, 2019; Hawes, 2020). These failures are a resounding rejection of the continued employment of traditional SDL methods. It is clear that we need alternative, and more reliable, privacy tools for the 2020 Census and beyond.

In pursuit of a modern paradigm for disclosure limitation, the Census Bureau endorsed differential privacy as the criterion to protect the public release of the 2020 Decennial Census data products. The Bureau openly engaged data users and sought constructive feedback when devising the new Disclosure Avoidance System (DAS). They launched a series of demonstration data product and codebase releases (U.S. Census Bureau, 2020a), and presented its design processes at numerous academic and professional society meetings, including the Joint Statistical Meeting, the 2020 National Academies of Sciences, Engineering, and Medicine (NASEM) Committee on National Statistics (CNSTAT) Workshop, and the 2019 Harvard Data Science Institute Conference in which I participated as a discussant. Reactions to this change from the academic data user communities were a passionate mix. Some cheered for the innovation, while others worried about the practical impact on the usability of differentially privatized releases. In keeping up with the inquiries and criticisms, the Census Bureau assembled and published data-quality metrics that were assessed repeatedly as the design of the 2020 DAS iterated (U.S. Census Bureau, 2020b). Through the process, the Bureau exhibited an unprecedented level of transparency and openness in conveying the design and the production of the novel disclosure control mechanism, publicizing the description of the TopDown Algorithm (Abowd et al., 2022) and the GitHub code base (2020 Census DAS Development Team, 2021). This knowledge makes a world of difference for census data users who need to analyze the privatized data releases and assess the validity and the quality of their work.

This article argues that *transparent* privacy enables principled statistical inference from privatized data releases. If a privacy mechanism is known, it can be incorporated as an integral part of a statistical model. Any additional uncertainty that the mechanism injects into the data can be accounted for properly. This is the most reliable way to ensure the correctness of the inferential claims produced from privatized data releases, when a calculated loss of statistical efficiency is present. For this reason, the publication of the probabilistic design of the privacy mechanism is crucial to maintaining a high usability of the privatized data product.

## 2. Differential Privacy Enables Transparency

Part of what contributed to the failure of the traditional disclosure limitation methods is that their justification appeals to intuition and obscurity, rather than explicit rules. If the released data are masked, coarsened, or perturbed from the confidential data, it seems natural to conclude that they are less informative, and consequently more 'private.' Traditional disclosure limitation mechanisms are obscure, in the sense that their design details are rarely released. For swapping-based methods, not only are the swap rates omitted, the attributes that have been swapped are often not disclosed (Oganian & Karr, 2006). As a consequence, an ordinary data user would not have the necessary information to replicate the mechanism, nor to assess their performance in protecting privacy. The effectiveness of obscure privacy mechanisms is difficult to quantify.

For data analysts who utilize data releases under traditional SDL to perform statistical tasks, the opaqueness of the privacy mechanism poses an additional threat to the validity of the resulting inference. A privacy mechanism, be it suppressive, perturbative, or otherwise, works by processing raw data and modifying their values to something that may be different from what has been observed. In doing so, the mechanism injects additional uncertainty in the released data, weakening the amount of statistical information contained in them. Uncertainty *per se* is not a problem; if anything, the discipline of statistics devotes itself to the study of uncertainty quantification. However, in order to properly attribute uncertainty where it is due, some minimal knowledge about its generative mechanism must be known. If the design of the privacy mechanism is kept opaque, our knowledge would be insufficient for producing reliable uncertainty estimates. The analyst might have no choice but to ignore the privacy mechanism imposed on the data, and might arrive at erroneous statistical conclusions.

Differential privacy conceptualizes privacy as the probabilistic knowledge to distinguish the identity of one individual respondent in the data set. The privacy guarantee is stated with respect to a random mechanism that imposes the privacy protection. Definition 1 presents the classic and most widely endorsed notion called $\epsilon$-*differential privacy*:

**Definition 1 ($\epsilon$-differential privacy; Dwork et al., 2006).** ***A mechanism $\tilde{S}:\mathbb{X}^n \to \mathbb{R}^p$ satisfies $\epsilon$-differential privacy, if for every pair of databases $\mathcal{D},\mathcal{D}' \in \mathbb{X}^n$ such that $\mathcal{D}$ and $\mathcal{D}'$ differ by one record, and every measurable set of outputs $A \in \mathscr{B}(\mathbb{R}^p)$, we have***

$$P\left(\tilde{S}(\mathcal{D}) \in A\right) \le e^{\epsilon} P\left(\tilde{S}(\mathcal{D}') \in A\right). \tag{2.1}$$

The positive quantity $\epsilon$, called the *privacy-loss budget* (PLB), enables the tuning, evaluation, and comparison of different mechanisms, all according to a standardized scale. In (2.1), the probability $P$ is taken with respect to the mechanism $\tilde{S}$, not with respect to the data $\mathcal{D}$.

As a formal approach to privacy, statistical disclosure limitation mechanisms compliant with differential privacy put forth two major advantages over their former counterparts. The first is *provability*, a mathematical formulation against which guarantees of privacy can be definitively verified as it is conceptualized. Definition 1 puts forth a concrete standard about whether, and by how much, any proposed mechanism can be deemed differentially private, as the probabilistic property of the mechanism is entirely encapsulated by $P$. As an example, we now understand that the classic randomized response mechanism (Warner, 1965), proposed decades before differential privacy, is in fact differentially private. Under randomized response, every respondent responds truthfully to a binary question with probability $p$, and with a random answer otherwise. That the randomized response mechanism is $\epsilon$-differentially private follows if $\epsilon$ is chosen such that $p = e^{\epsilon}/\left(1+e^{\epsilon}\right)$ (see, e.g. (Dwork & Roth, 2014, Theorem 3.14)). With provability, anyone can design new mechanisms with privacy guarantees under an explicit rule, as well as to verify whether a publicized privacy mechanism lives up to its guarantee.

The second major advantage of differential privacy, which this article underscores, is *transparency*. Differential privacy allows for the full, public specification of the privacy mechanism without sabotaging the privacy guarantee. The data curator has the freedom to disseminate the design of the mechanism, allowing the data users to utilize it and to critique it, without compromising the effectiveness of the privacy protection. The concept of transparency that concerns this article will be made precise in Section 4. As a example, below is one of the earliest proposed mechanisms that satisfies differential privacy:

**Definition 2** (Laplace mechanism; Dwork et al., 2006). ***Given a confidential database $\mathcal{D} \in \mathbb{X}^n$, a deterministic query function $S: \mathbb{X}^n \to \mathbb{R}^p$ and its global sensitivity $\Delta\left(S\right)$, the $\epsilon$-differentially private Laplace mechanism is***

$$\tilde{S}\left(\mathcal{D}\right) = S\left(\mathcal{D}\right) + \left(U_1, \ldots, U_p\right),$$

***where $U_i$'s are real-valued i.i.d. random variables with $\mathbb{E}(U_i) = 0$ and probability density function***

$$f\left(u\right) \propto e^{-\frac{\epsilon|u|}{\Delta(S)}}. \tag{2.2}$$

The omitted proportionality constant in (2.2) is equal to $\epsilon/2\Delta\left(S\right)$, ensuring that the density $f$ integrates to one. The global sensitivity $\Delta(S)$ measures the maximal extent to which the deterministic query function changes in response to the perturbation of one record in the database. For counting queries operating on binary databases, such as population counts, $\Delta(S) = 1$. To note is that in Definition 2, both the deterministic query $S$ and the probability distribution of the noise terms $U_i$'s are fully known. Anyone can implement the privacy algorithm on a database of the same form as $\mathcal{D}$.

We note that differentially private mechanisms *compose* their privacy losses nicely. At a basic level, two separately released differentially private data products, incurring PLBs of $\epsilon_1$ and $\epsilon_2$ respectively, incur no more than a total PLB of $\left(\epsilon_1 + \epsilon_2\right)$ when combined (Dwork & Roth, 2014, Theorem 3.14). Superior composition, reflecting a more efficient use of PLBs, can be achieved with the clever design of privacy mechanisms. The composition property provides assurance to the data curator that when releasing multiple data products over time, the total privacy loss can be controlled and budgeted ahead of time.

The preliminary versions of the 2020 Census DAS utilizes the integer counterpart to the Laplace mechanism, called the double geometric mechanism (Fioretto et al., 2021; Ghosh et al., 2012). The mechanism possesses the same additive form as the Laplace mechanism, but instead of real-valued noise $U_i$'s, it uses integer-valued ones whose probability mass function has the same form as (2.2) with the proportionality constant equal to $(1-e^{-\epsilon})/(1+e^{-\epsilon})$. The production implementation of the DAS, used for the P.L. 94-171 public release (U.S. Census Bureau, 2021a) and the 2021-06-08 vintage demonstration files (Van Riper et al., 2020), appeals to a variant privacy definition called the *zero-concentrated differential privacy* (Dwork & Rothblum, 2016).[1] It employs additive noise with discrete Gaussian distributions according to a detailed PLB schedule (U.S. Census Bureau, 2021c). While all mechanisms discussed above employ additive errors, differential privacy mechanisms in general need not be additive. Non-additive examples commonly used in the private computation of complex queries include the exponential mechanism (McSherry & Talwar, 2007), objective perturbation (Kifer et al., 2012), and the $K$-norm gradient mechanism (Reimherr & Awan, 2019). In what follows, we elaborate on the importance of transparent privacy from the statistical point of view.

# 3. What Can Go Wrong With Obscure Privacy

Data privatization constitutes a phase in data processing which succeeds data collection and precedes data release. When conducting statistical analysis on processed data, misleading answers await if the analyst ignores the phases of data processing and the consequences they impose.

We use an example of simple linear regression to illustrate how obscure privacy can be misleading. Regression models occupy a central role in many statistical analysis routines, for they can be thought of as a first-order approximation to any functional relationship between two or more quantities. Let $(x_i, y_i)$ be a pair of quantities measured across a collection of geographic regions indexed by $i = 1, \ldots, n$. Examples of $x_i$ and $y_i$ may be counts of population of certain demographic characteristics within each census block of a state, households of certain types, economic characteristics of the region (businesses, revenue, and taxation; see, e.g., Barrientos et al., 2021), and so on. Suppose the familiar simple regression model is applied:

$$y_i = \beta_0 + \beta_1 x_i + e_i, \tag{3.1}$$

where the $e_i$'s are independently and identically distributed idiosyncratic errors with mean zero and variance $\sigma^2$, typically following the normal distribution. Usual estimation techniques for $\beta_0$ and $\beta_1$, such as ordinary least squares or maximum likelihood, produce unbiased point estimators. For the slope, $\hat{\beta}_1 = \sum_{i=1}^{n}(x_i - \bar{x})(y_i - \bar{y}) / \sum_{i=1}^{n}(x_i - \bar{x})(x_i - \bar{x})$ where $\mathbb{E}\left(\hat{\beta}_1\right) = \beta_1$, and for the intercept, $\hat{\beta}_0 = \bar{y} - \hat{\beta}_1\bar{x}$ where $\mathbb{E}\left(\hat{\beta}_0\right) = \beta_0$, where expectations are taken with respect to variabilities in the error terms. Both estimators also enjoy consistency when the regressor $x_i$'s are random, that is, $\hat{\beta}_1 \to \beta_1$ and $\hat{\beta}_0 \to \beta_0$, indicating convergence in probability as the sample size $n$ approaches infinity. The consistency of $(\hat{\beta}_0, \hat{\beta}_1)$ is reasonably robust against mild heteroskedasticity of the idiosyncratic errors.

Since $x_i$ and $y_i$ contain information about persons and businesses that may be deemed confidential, suppose they are privatized before release using standard additive differential privacy mechanisms. Their privatized versions $(\tilde{x}_i, \tilde{y}_i)$ are respectively

$$\tilde{x}_i = x_i + u_i, \qquad \tilde{y}_i = y_i + v_i. \tag{3.2}$$

The $u_i$'s and $v_i$'s can be chosen according to the Laplace mechanism or the double geometric mechanism following Definition 2, with suitable scale parameters such that $(\tilde{x}_i, \tilde{y}_i)$ are compliant with $\epsilon$-differential privacy, and accounting for the sequential composition of the $\tilde{x}_i$'s and $\tilde{y}_i$'s. We denote the variances of $u_i$ and $v_i$ as $\sigma_u^2$ and $\sigma_v^2$, respectively.[2] As the privacy budget allocated to either statistic decreases, the privacy error variance increases and more privacy is achieved, and vice versa.

Suppose the analyst is supplied the privatized statistics $(\tilde{x}_i, \tilde{y}_i)$, but is not told how they are generated based on the confidential statistics $(x_i, y_i)$. That is, (3.2) is entirely unknown to her. In this situation, there is no obvious way for her to proceed, other than to ignore the privacy mechanism and run the regression analysis by treating the privatized $(\tilde{x}_i, \tilde{y}_i)$ as if they're the confidential values. If so, the analyst would effectively perform parameter estimation for a different, *naïve* linear model

$$\tilde{y}_i = b_0 + b_1 \tilde{x}_i + \tilde{e}_i. \tag{3.3}$$

Unfortunately, no matter which computational procedure one uses, the point estimates obtained from fitting (3.3) are no longer unbiased nor consistent for $\beta_0$ and $\beta_1$ as in the original model of (3.1). Both naïve estimators, call them $\hat{b}_0$ and $\hat{b}_1$, are complex functions that convolute the confidential data, idiosyncratic errors, and privacy errors. When the regressor $x_i$'s are random realizations from an underlying infinite population, the bias inherent to the naïve estimators does not diminish even if the sample size approaches infinity. More precisely, we have that the naïve slope estimator

$$\hat{b}_1 \to \frac{\mathbb{V}(x)}{\mathbb{V}(x)+\sigma_u^2}\beta_1, \tag{3.4}$$

and the naïve intercept estimator

$$\hat{b}_0 \to \beta_0 + \left(1 - \frac{\mathbb{V}(x)}{\mathbb{V}(x)+\sigma_u^2}\right)\mathbb{E}(x)\,\beta_1,$$

where $\mathbb{E}(x)$ and $\mathbb{V}(x)$ are the population-level mean and variance of $x$ for which the observed sample is representative. The ratio $\mathbb{V}(x)\,/\left(\mathbb{V}(x) + \sigma_u^2\right)$ displays the extent of inconsistency of $\hat{b}_1$ as a function of the population variance and the privacy error variance of $x$. We see that if the independent variable is not already centralized, $\hat{b}_0$ exhibits a bias whose magnitude is influenced by both the average magnitude of $x$, as well as the amount of privacy protection for $x$. In addition, the residual variance from the naïve linear model (3.3) is also inflated, with

$$\mathbb{V}(\tilde{y} \mid \tilde{x}) = \sigma^2 + \beta_1^2 \sigma_u^2 + \sigma_v^2, \tag{3.5}$$

which is strictly larger than $\sigma^2$, the usual residual variance from the correct linear model (3.1). If the independent variable $x_i$'s are treated as fixed instead, an exact finite-sample characterization of the naïve estimators $\hat{b}_0$ and $\hat{b}_1$ are difficult to obtain. Appendix A presents the distribution limits for the slope estimator as a function of the scales of the privacy errors and the regression errors, and showcases how the coverage probabilities deteriorate as the privacy errors increase.

We use a small sample simulation study ($n = 10$) to illustrate the pitfall with obscure privacy. Assume that the confidential data follows the generative process of (3.1), with $x_i \sim Pois(10)$ i.i.d., $\sigma = 5$, and the true parameter values $(\beta_0, \beta_1) = (-5, 4)$. The privatized data $(\tilde{x}_i, \tilde{y}_i)$ are subsequently generated according to the additive privacy mechanism of (3.2), where $u_i, v_i \sim \text{Laplace}(\epsilon^{-1})$, with a PLB of $\epsilon = 0.25$. The three panels of Figure 1 depict different statistical inference—both right and wrong types—that correspond to three scenarios in this example. When no privacy protection is enforced, a 95% confidence ellipse for $(\beta_0, \beta_1)$ from the simple linear regression should cover the true parameter values (represented by the orange square) at approximately the nominal coverage rate, a high probability of 95%. The left panel displays one such confidence ellipse in blue. When privacy protection is in place, directly fitting the linear regression model on $(\tilde{x}, \tilde{y})$ may result in misleading inference, as can be seen from the naïve 95% confidence ellipses (gray) in the middle panel, all derived from privatized versions of the same confidential data set, repeatedly miss their mark as they rarely cover the true value. We witness the biasing behavior precisely as established: the slope $\beta_1$ is underestimated, displaying a systematic shrinking toward zero, whereas the true value of $\beta_0$ is overestimated, with $\beta_1 > 0$ and $\mathbb{E}(x) = 10 > 0$. In contrast, the green ellipses in the right panel, each representing an approximate 95% confidence region, are based on the correct analysis on privatized data accounting for the privacy mechanism (to be discussed in Section 4). They better recover the location of the true parameters, and display larger associated inferential uncertainty.

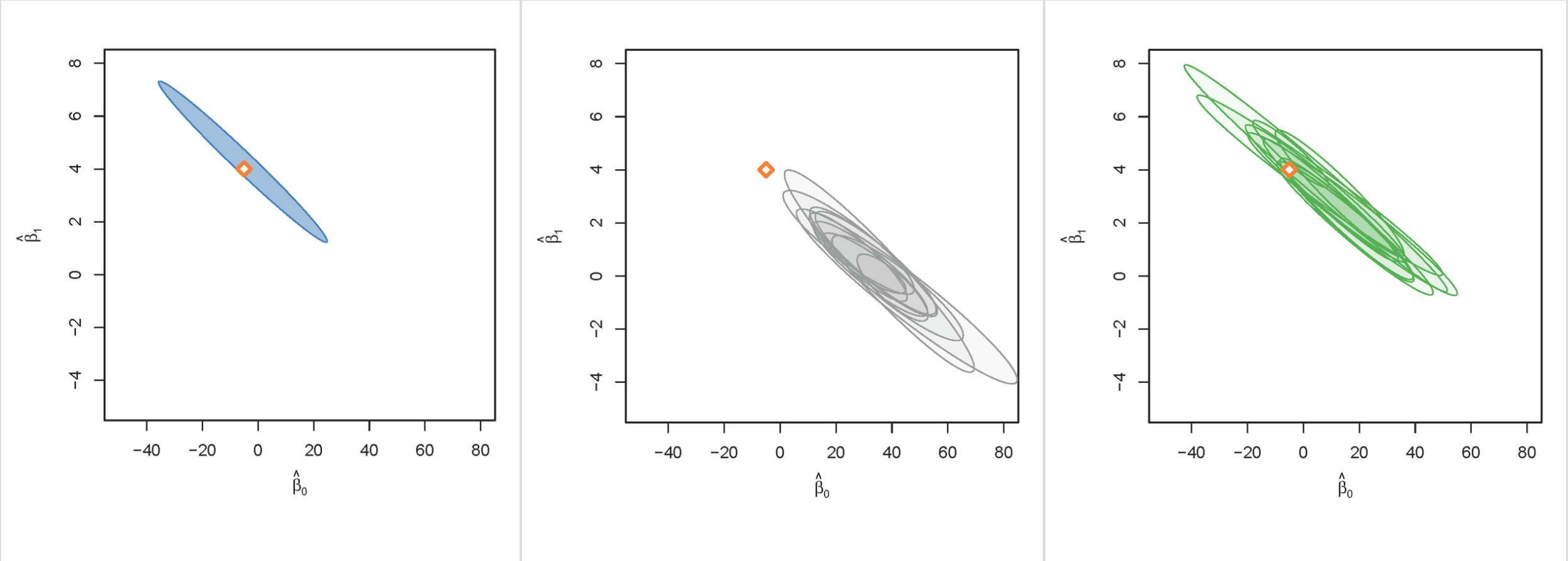

**Figure 1.** 95% joint confidence regions for $(\beta_0, \beta_1)$ from linear regression (3.1). Left: original data $(x, y)$ simulated according to (3.1). Middle: naïve linear regression (3.3) on $n = 10$ pairs of simulated privatized data $(\tilde{x}, \tilde{y})$ from the Laplace mechanism (3.2) with PLB of $\epsilon = 0.25$. Right: the correct model following (4.2) implemented using Monte Carlo expectation maximization on the same sets of private data. Concentration ellipses are large-sample approximate 95% confidence regions based on estimated Fisher information at the maximum likelihood estimate. The orange square represents the ground truth $(\beta_0, \beta_1) = (-5, 4)$.

The troubling consequence of ignoring the privacy mechanism is not new to statisticians. The naïve regression analysis of privatized data generalizes a well-known scenario in the measurement error literature, called the *classic* measurement error model. The notable biasing effect on the naïve estimator $\hat{b}_1$ created by the additive noise (3.2) in the independent variable $x$ is termed *attenuation*. The bias causes a “double whammy” (Carroll et al., 2006, Chapter 15) on the quality of the resulting inference, because one is misled in terms of both the location of the true parameter, and the extent of uncertainty associated with the estimators, as seen from the erroneous coverage probability within its asymptotic sampling distributional limits. In linear models, additive measurement errors in the dependent variable $y$ is generally considered less damaging, because if the errors are independent, unbiased, and present in the dependent variable only, the model fit remains unbiased (Carroll et al., 2006, Chapter 15), hence such errors are often ignored or treated as a component to the idiosyncratic regression errors. However, they would still increase the variability of the fitted model and decrease the statistical power in detecting an otherwise significant effect. Consequently, they may still negatively affect the quality of any naïve model fitting on privatized data, both by changing the effective nominal coverage rate of the large sample distribution limits (see Appendix A for details), and by increasing uncertainty of the fitted model according to (3.5).

From the additive mechanism in Definition 2, we see that the noise term $u_i$ is a symmetric, zero-mean random variable. This means that the privacy mechanism is *unbiased* for its underlying query: it has the exact same chance to inflate or deflate it in either direction by the same magnitude. How can an unbiased privatization algorithm, followed by an unbiased statistical procedure (i.e., the simple linear regression), results in biased statistical estimates? The issue is that while the privatized data $\tilde{S}$ is unbiased for the confidential data $S$, if the estimator we use is a *nonlinear* function of $S$, it may no longer retain unbiasedness if $S$ were perturbed. The regression coefficients $\hat{\beta}_0$ and $\hat{\beta}_1$ are nonlinear estimators. Specifically, $\hat{\beta}_1$ is a ratio estimator, and $\hat{\beta}_0$ depends on $\hat{\beta}_1$ as a building block. In general, the validity of ratio estimators are particularly susceptible to minor instabilities in its denominator. Replacing confidential statistics with their unbiased privatized releases may not be an innocent move, if the downstream analysis calls for nonlinear estimators that cannot preserve unbiasedness.

In the universe of statistical analysis, nonlinear estimators are the rule, not the exception. Many descriptive and summary statistics involve nonlinear operations such as squaring or dividing—think variances, proportions, and other complex indices[3]—which don’t fare well with additive noise. Ratio estimators, or estimators that involve random quantities in their denominators, can suffer from high variability if the randomness is high.

Therefore, many important use cases of the census releases, as well as the assessment of the impact due to privacy, could benefit from additional uncertainty quantification. As an example, Asquith et al. (2022) evaluate a preliminary version of the 2020 Census DAS using a set of segregation metrics as the benchmark statistics and compare its effect when applied to the 1940 full-count census microdata. One of the evaluation metrics is the *index of dissimilarity* per county (Iceland et al., 2002):

$$d = \frac{1}{2} \sum_{i=1}^{n} \left| \frac{w_i}{w_{\text{cty}}} - \frac{b_i}{b_{\text{cty}}} \right|, \tag{3.6}$$

where $w_i$ and $b_i$ are respectively the White and the African American populations of tract $i$ of the county, and $w_{\text{cty}}$ and $b_{\text{cty}}$ those of the entire county. All of these quantities are subject to privacy protection, and one run of the DAS creates a version of $\{\tilde{w}_i, \tilde{b}_i, \tilde{w}_{\text{cty}}, \tilde{b}_{\text{cty}}\}$, each infused with Laplace-like noise.

If we were to repeatedly create privatized demonstration data sets from the DAS, and calculate the dissimilarity index each time by naïvely replacing all quantities in (3.6) with their privatized counterparts, we will witness variability in the value $d$. Since $d$ is a ratio estimator, its value may exhibit a large departure from the confidential true value, particularly when the denominator is small, such as when a county has a small population, or is predominantly White or non-White. Since every DAS output is uniquely realized by a single draw from its probabilistic privacy mechanism, the value $d$ calculated based on a particular run of the DAS will exhibit a difference from its confidential (or true) value.[4] The difference will be unknown, but can be described by the known properties of the privacy mechanism. It is important to recognize the probabilistic nature of the statistics calculated from privatized data, and interpret them alongside appropriate uncertainty quantification, which itself is a reflection of data quality.

Privacy adds an extra layer of uncertainty to the generative process of the published data, just as any data-processing procedures such as cleaning, smoothing, or missing data imputation. We risk obtaining misguided inference whenever blindly fitting a favorite confidential data model on privatized data without acknowledging the privatization process, for the same reason we would be misguided by not accounting for the effect of data processing. To better understand the inferential implication of privacy and obtain utility-oriented assessments, privacy shall be viewed as a controllable source of total survey error, an approach that is again made feasible by the transparency of the privatization procedure. We return to this subject in Section 5.

# 4. Principled Analysis With Transparent Privacy

The misleading analysis presented in Section 3 is not the fault of differential privacy, nor of linear regression or other means of statistical modeling. Rather, obscure privacy mechanisms prevent us from performing the right analysis. Any statistical model, however adequate in describing the probabilistic regularities in the confidential data, will generally be inadequate when naïvely applied to the privatized data.

## 4.1. Accounting for the Privacy Mechanism

To correctly account for the privacy mechanism, statistical models designed for confidential data need to be *augmented* to include the additional layer of uncertainty due to privacy. In our example, the simple linear model of (3.1) is the true generative model for the confidential statistics $(x, y)$. Together with the privacy mechanism in (3.2), the implied true generative model for the privatized statistics $(\tilde{x}, \tilde{y})$ can be written as

$$\tilde{y}_i + v_i = \beta_0 + \beta_1 \left(\tilde{x}_i + u_i\right) + e_i, \tag{4.1}$$

where $u_i, v_i$ are additive privacy errors and $e_i$ the idiosyncratic regression error. Thus, with the original linear model (3.1) being the correct model for $(x, y)$, it follows that the augmented model (4.1) is the correct model for describing the stochastic relationship between $(\tilde{x}, \tilde{y})$. On the other hand, unless all $u_i$'s and $v_i$'s are exactly zero, that is, no privacy protection is effectively performed for both $x$ and $y$, the naïve model in (3.3) is erroneous and incommensurable with the augmented model in (4.1).

If a statistical model is of high quality, or more precisely *self-efficient* (Meng, 1994; Xie & Meng, 2017),[5] its inference based on the privatized data should typically bear more uncertainty compared to that based on the confidential data. The increase in uncertainty is attributable to the privacy mechanism. Therefore, uncertainty quantification is of particular importance when it comes to analyzing privatized data. But drawing statistically valid inference from privatized data is not as simple as increasing the nominal coverage probability of confidence or credible regions from the old analysis. As we have seen, fitting the naïve linear model on differentially privatized data creates a 'double whammy' due to both a biased estimator and incorrectly quantified estimation uncertainty. The right analysis hinges on incorporating the probabilistic privacy mechanism into the model itself. This ensures that we capture uncertainty stemming from any potential systematic bias displayed by the estimator due to noise injection, as well as a sheer loss of precision due to diminished informativeness of the data.

For data users who currently employ analysis protocols designed without private data in mind, this suggests that modification needs to be made to their favorite tools. That sounds like an incredibly daunting task. However, on a conceptual level, what needs to be done is quite simple. We present a general recipe for the vast class of statistical methods with either a likelihood or a Bayesian justification.

Let $\beta$ denote the estimand of interest. For randomization-based inference common to the literatures of survey and experimental design, this estimand may be expressed as a function of the confidential database: $\beta = \beta(\mathcal{D})$. In model-based inference, $\beta$ may be the finite- or infinite-dimensional parameter that governs the distribution of $\mathcal{D}$. Let $\mathcal{L}$ be the original likelihood for $\beta$ based on the confidential data $s$, representing the currently employed, or ideal, statistical model for analyzing data that is not subject to privacy protection. Choices for $\beta$ and $\mathcal{L}$ are both made by the data analyst.

Let $p_\xi\left(\tilde{s} \mid s\right)$ be the conditional probability distribution of the privatized data $\tilde{s}$ given $s$, as induced by the privacy mechanism chosen by the data curator. The subscript $\xi$ encompasses all tuning parameters of the mechanism, as well as any auxiliary information that is used during the privatization process. Note that the mechanism $p_\xi$ need not be a *differentially* private mechanism: it may be a traditional SDL mechanism, or any other mechanism that the data curator chooses to impose on the confidential data, probabilistic or otherwise. As an example, $p_\xi$ may stand for the class of swapping methods, in which case $\xi$ encodes the swap rates and the list of the variables being swapped. If $p_\xi$ is induced by the Laplace mechanism in Definition 2, then $\xi$ stands for the class of product Laplace densities centered at $s$, and $\xi$ its scale parameter which, if set to $\Delta(S)/\epsilon$, qualifies $p_\xi$ as an $\epsilon$-differentially private mechanism.

**Definition 3** (Transparent privacy). ***A privacy mechanism is said to be transparent if*** $p_\xi\left(\cdot \mid \cdot\right)$ ***, the conditional probability distribution it induces given the confidential data, is known to the user of the privatized data, including both the functional form of*** $p_\xi$ ***and the specific value of*** $\xi$ ***employed.***

When the privacy mechanism is transparent, we can write down the *observed*, or *marginal*, likelihood function for $\beta$ based on the observed $\tilde{s}$ (Williams & Mcsherry, 2010):

$$\mathcal{L}_\xi\left(\beta; \tilde{s}\right) = \int p_\xi\left(\tilde{s} \mid s\right) \mathcal{L}\left(\beta; s\right) ds, \tag{4.2}$$

with the notation $\mathcal{L}_\xi$ highlighting the fact that it is a weighted version of the original likelihood $\mathcal{L}$ according to the privacy mechanism $p_\xi$. The integral expression of (4.2) is reminiscent of the missing data formulation for parameter estimation (Little & Rubin, 2014). The observed data is the privatized data $\tilde{s}$, and the missing data is the confidential data $s$, with the two of them associated by the privacy mechanism $p_\xi$ analogous to the missingness mechanism. All information that can be objectively learned about the parameter of interest $\beta$ has to be based on the observed data alone, averaging out the uncertainties in the missing data. In the regression example, the observed likelihood is precisely the joint probability distribution of $\left(\tilde{x}_i, \tilde{y}_i\right)$ according to the implied true model (4.1), governed by the parameters $\beta_0$ and $\beta_1$, with sampling variability derived from that of the idiosyncratic errors $e_i$ as well as privacy errors $u_i$ and $v_i$. All modes of statistical inference congruent with the original data likelihood $\mathcal{L}$, including frequentist procedures that can be embedded into $\mathcal{L}$ as well as Bayesian models based on $\mathcal{L}$, would have adequately accounted for the privacy mechanism by respecting (4.2). Furthermore, for a Bayesian analyst who employs a prior distribution for $\beta$, denoted as $\pi_0$, her posterior distribution now becomes

$$\pi_\xi\left(\beta \mid \tilde{s}\right) = c_\xi \pi_0\left(\beta\right) \mathcal{L}_\xi\left(\beta; \tilde{s}\right), \tag{4.3}$$

where the proportionality constant $c_\xi$, free of the parameter $\beta$, ensures that the posterior integrates to one.

## 4.2. The Necessity of Transparent Privacy

The marginal likelihood for $\beta$ in (4.2) highlights why transparency allows data users to achieve inferential validity for their question of interest from privatized data. To compute quantities based on this likelihood, one must know not only the original statistical model $\mathcal{L}$ but also the privacy mechanism $p_\xi$, including its parameter $\xi$. We formalize the crucial importance of transparent privacy in ensuring inferential validity.

**Theorem 1 (Necessity of transparent privacy).** ***Let $\beta \in \mathbb{R}^d$ be a continuous parameter and $h(\beta)$ a bounded Borel-measureable function for which inference is sought. The observed data $\tilde{s}$ is privatized with the mechanism $p_\xi(\cdot \mid s)$, and the analyst supposes the mechanism to be $q(\cdot \mid s)$. Then for all likelihood specifications $\mathcal{L}$ with base measure $\nu$, observed data $\tilde{s}$ and choice of $h$, the analyst recovers the correct posterior expectation for $h(\beta)$, i.e.***

$$E_q(h(\beta) \mid \tilde{s}) = E_\xi(h(\beta) \mid \tilde{s}) \tag{4.4}$$

***if only if $p_\xi(\cdot \mid s) = q(\cdot \mid s)$ for $\nu$-almost all $s$.***

*Proof.* The 'if' part of the theorem is trivial. For the 'only if' part, note that (4.4) is the same as the requirement of weak equivalence between the true posterior $\pi_\xi(\beta \mid \tilde{s})$ in (4.3) and the analyst's supposed posterior:

$$\pi_q(\beta \mid \tilde{s}) := c_q \pi_0(\beta) \int q(\tilde{s} \mid s) \mathcal{L}(\beta; s)\, ds,$$

where the proportionality constant $c_q$, free of $\beta$, ensures that the density $\pi_q$ integrates to one. This in turn requires for any given $\tilde{s}$ and the constant $c = c_\xi / c_q > 0$,

$$E(q(\tilde{s} \mid s) - c p_\xi(\tilde{s} \mid s) \mid \beta) = 0$$

for $\beta \in \mathbb{R}^d$ almost everywhere, where the expectation above is taken with respect to the likelihood $\mathcal{L}$. Since $\mathcal{L}$ is chosen by the analyst but $p_\xi$ is not, this implies that she must also choose $q$ so that $q(\tilde{s} \mid s) - c p_\xi(\tilde{s} \mid s) = 0$ for all $s$ except on a set of measure zero relative to $\nu$. Furthermore, since $\int q(a \mid s)\, da = \int p_\xi(a \mid s)\, da = 1$ for every $s$, we must have $c = 1$, thus $p_\xi(\cdot \mid s) = q(\cdot \mid s)$ as desired.

What Theorem 1 says is that, if we conceive the statistical validity of an analysis as its ability to yield the same expected answer as that implied by the correct model (that is, by properly accounting for the privatization mechanism) for a wide range of questions (reflected by the free choice of $h$), then the only way to ensure statistical validity is to grant the analyst full knowledge of the probabilistic characteristics of the privatization mechanism.

As discussed in Section 1, traditional SDL techniques such as suppression, deidentification, and swapping rely fundamentally on procedural secrecy. While each of these methods admits a precise characterization $p_\xi$, such information—in particular, the production settings of $\xi$—is intentionally kept out of public view. The lack of transparency with traditional SDL mechanisms hinders the possibility to draw principled and statistically valid inference from data products they produce.

Scholars in the SDL literature advocate for transparent privacy for more than one good reason. With a rearrangement of terms, the posterior in (4.3) can also be written as (details in Appendix B)

$$\pi_\xi(\beta \mid \tilde{s}) = \int \pi(\beta \mid s)\, \pi_\xi(s \mid \tilde{s})\, ds, \tag{4.5}$$

where $\pi(\beta \mid s)$ is the posterior model for the confidential $s$, and $\pi_\xi(s \mid \tilde{s})$ the posterior predictive distribution of the confidential $s$ based on the privatized $\tilde{s}$, again with its dependence on the privacy mechanism $p_\xi$ highlighted in the subscript. This representation of the posterior resembles the theory of multiple imputation (Rubin, 1996), which lies at the theoretical foundation of the synthetic data approach to SDL (Raghunathan et al., 2003; Rubin, 1993). What (4.5) illustrates is an alternative viewpoint on private data analysis. The correct Bayesian analysis can be constructed as a mixture of naïve analyses based on the agent's best knowledge of the confidential data, where this best knowledge is instructed by the privatized data, the prior, as well as the transparent privatization procedure. Under this view, the transparency of the privacy mechanism again becomes a crucial ingredient to the *congeniality* (Meng, 1994; Xie & Meng, 2017) between the imputer's model and the analyst's model, ensuring the quality of inference the analyst can obtain. Karr and Reiter (2014, p.284) call the Bayesian formulation (4.5) the "SDL of the future," emphasizing the insurmountable computational challenge the analyst would otherwise need to face without knowing the term $\pi_\xi(s \mid \tilde{s})$. With transparency of $p_\xi$ at hand, the future is in sight.

Transparent privacy mechanisms merit another important quality, namely *parameter distinctiveness*, or a priori parameter independence, from both the generative model of the true confidential data as well as any descriptive model the analyst wishes to impose on it. Parameter distinctiveness always holds since the entire privacy mechanism, all within control of the curator, is fully announced hence has no hidden dependence on the unknown inferential parameter $\beta$ through means beyond the confidential data $s$. In the missing data literature, parameter distinctiveness is a prerequisite of the missing data mechanism to give way for simplifying assumptions, such as missing completely at random (MCAR) and missing at random (MAR; Rubin, 1976), allowing for the missingness model to sever any dependence on the unobserved data.[6] In the privacy context, parameter distinctiveness ensures that the privacy mechanism does not interact with any modeling decision imposed on the confidential data. It is the reason why the true observed likelihood $\mathcal{L}_\xi$ in (4.2) involves merely two terms, $p_\xi$ and $\mathcal{L}$, whose product constitutes the implied joint model for the complete data $(s, \tilde{s})$ for every choice of $\mathcal{L}$. This may result in potentially vast simplification in many cases of downstream analysis. The practical benefit of parameter distinctiveness of the privacy mechanism is predicated on its transparency, for unless a mechanism is *known* (Abowd & Schmutte, 2016), none of its properties can be verified nor put into action with confidence.

While conceptually simple, carrying through the correct calculation can be computationally demanding. The integral in (4.2) may easily become intractable if the statistical model is complex, if the confidential data is high-dimensional (as is the case with the census tabulations), or if a combination of both holds true. The challenge is amplified by the fact that the two components of the integral are generally not in conjugate forms. While the privacy mechanism $p_\xi$ is determined by the data curator, the statistical model $\mathcal{L}$ is chosen by the data analyst, and the two parties typically do not consult each other in making their respective choices. Even for the simplest models, such as the running linear regression example, we cannot expect (4.2) to possess an analytical expression.

To answer to the demand for statistically valid inference procedures based on privatized data, Gong (2019) discusses two sets of computational frameworks to handle independently and arbitrarily specified privacy mechanisms and statistical models. For exact likelihood inference, the integration in (4.2) can be performed using Monte Carlo expectation maximization (MCEM), designed for the presence of latent variables or partially missing data and equipped with a general-purpose importance sampling strategy at its core. Exact Bayesian inference according to (4.3) can be achieved with, somewhat surprisingly, an approximate Bayesian computation (ABC) algorithm. The tuning parameters of the ABC algorithm usually control the level of approximation in exchange for Monte Carlo efficiency, or computational feasibility in complex models. In the case of privacy, the tuning parameters are set to reflect the privacy mechanism, in such a way that the algorithm outputs exact draws from the desired Bayesian posterior for any proper prior specification. I have explained this phenomenon with a catchy phrase: approximate computation on exact data is exact computation on approximate data. Private data is approximate data, and its inexact nature can be leveraged to our benefit, if the privatization procedure becomes correctly aligned with the necessary approximation that brings computational feasibility.

To continue the illustration with our running example, the MCEM algorithm is implemented to draw maximum likelihood inference for the $\beta$'s using privatized data. The right panel of Figure 1 depicts 95% approximate confidence regions (green) for the regression coefficients based on simulated privatized data sets $(\tilde{x}, \tilde{y})$ of size $n = 10$. The confidence ellipses are derived using a normal approximation to the likelihood at the maximum likelihood estimate, with covariance equal to the inverse observed Fisher information. Details of the algorithm can be found in Appendix C. We see that the actual inferential uncertainty for both $\beta_0$ and $\beta_1$ are inflated compared to inference on confidential data as in the left panel, but in contrast to the naïve analysis in the middle panel, most of these green ellipses cover the ground truth despite a loss of precision. The inference they represent adequately reflects the amount of uncertainty present in the privatized data.

## 5. Privacy as a Transparent Source of Total Survey Error

In introductory probability and survey sampling classrooms, the concept of a census is frequently invoked as a pedagogical reference, often with the U.S. Decennial Census as a prototype. The teacher would contrast statistical inference from a probabilistic sampling scheme with directly observing a quantity from the census, regarding the latter as the gold standard, if not the ground truth. This narrative may have left many quantitative researchers with the impression that the census is always comprehensive and accurate. The reality, however, invariably departs from this ideal. The census is a survey, and is subject to many kinds of errors and uncertainties, as are all surveys. As do coverage bias, nonresponse, erroneous and edited inputs, statistical disclosure limitation introduces a source of uncertainty into the survey, albeit unique in nature.

To assess the quality of the end data product, and to improve it to the extent possible, we construe privacy as one of the several interrelated contributors to total survey error (TSE; Groves, 2005). Errors due to privacy make up a source of nonsampling survey error (Biemer, 2010). Additive mechanisms create privacy errors that bear a structural resemblance with measurement errors (Reiter, 2019). What makes privacy errors easier to deal with than other sources of survey error, at least theoretically, is that their generative process is verifiable and manipulable. Under central models of differential privacy, the process is within the control of the curator, and under local models (i.e., the responses are privatized as they leave the respondent) it is defined by explicit protocols. Transparency brings several notable advantages to the game. Privacy errors are known to enjoy desirable properties such as simple and tractable probability distributions, statistical independence among the error terms, as well as between the errors and the underlying confidential data (i.e., parameter distinctiveness). These properties may be assumptions for measurement errors, but they are known to hold true for privacy errors. In the classic measurement error setting, the error variance needs to be estimated. In contrast, the theoretical variance of all the additive privacy mechanisms are known and public. The structural similarity between privacy errors and measurement errors allows for the straightforward adaptation of existing tools for measurement error modeling, including regression calibration and simulation extrapolation, which perform well for a wide class of generalized linear models. Other approaches that aim to remedy the effect of both missing data and measurement errors can be modified to include privacy errors (Blackwell et al., 2017a; Blackwell et al., 2017b; Kim et al., 2014; Kim et al., 2015). Most recently, steps are being taken to develop methods for direct bias correction in the regression context (Evans & King, n.d.).

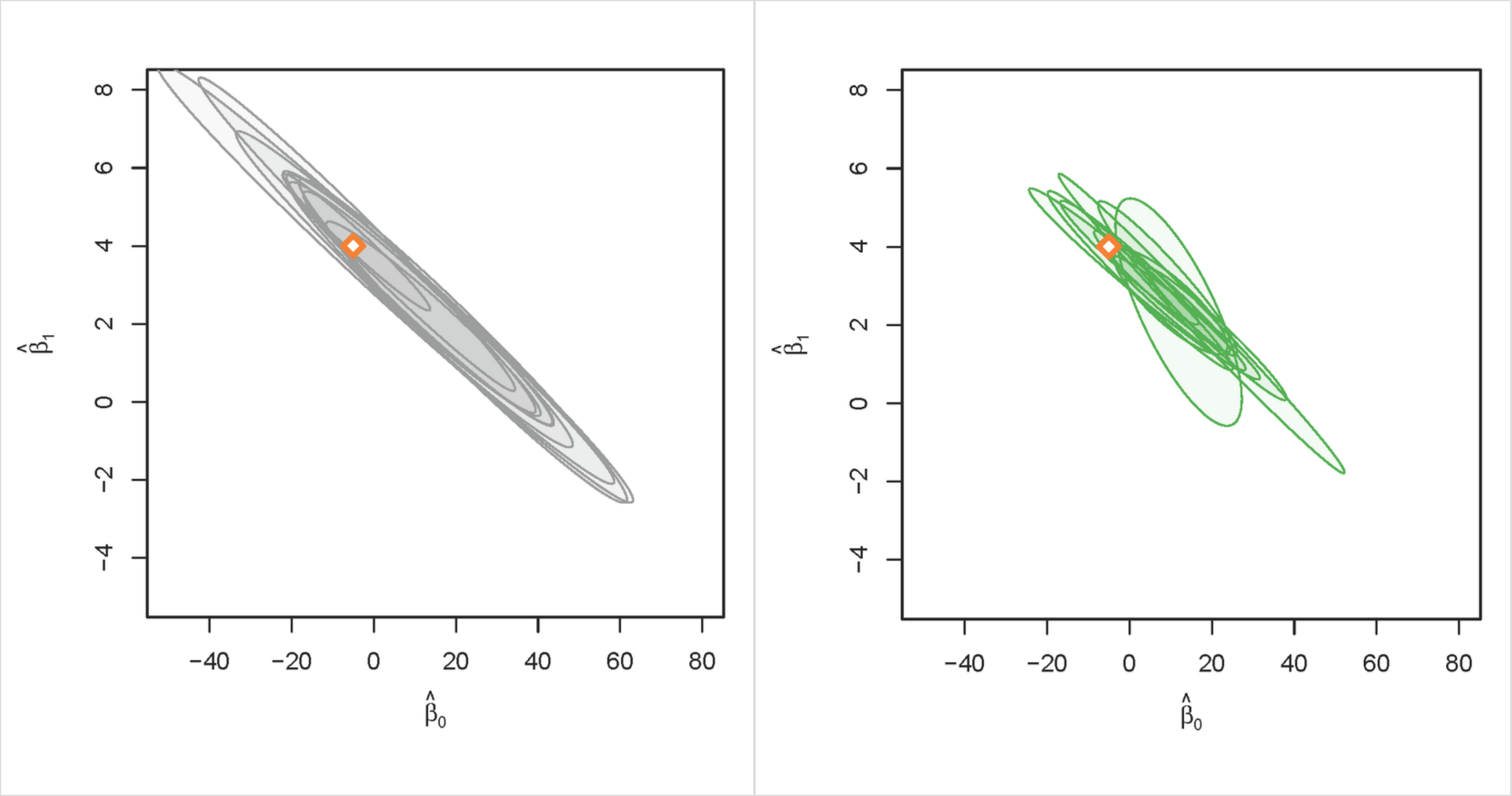

**Figure 2.** 95% joint confidence regions for $(\beta_0, \beta_1)$ derived from the same set of linear regression analyses on privatized data as depicted in Figure 1, but with $\epsilon = 1$, a four-fold privacy-loss budget increase. While the correct, Monte Carlo expectation maximization–based analysis (right) remains valid, the accuracy of the naïve analysis (left) is greatly improved (compared to the middle panel of Figure 1), at the expense of a weaker privacy guarantee from the data.

We emphasize that the transparency of the privacy mechanism is crucial to the understanding, quantification, and control of its impact on the quality of the resulting data product from a total survey of error approach. As noted in Karr (2017), traditional disclosure limitation methods often passively interact with other data-processing and error-reduction procedures commonly applied to surveys, and the effect of such interactions is often subtle. Due to the artificial nature of all privacy mechanisms, any interaction between the privacy errors can be explicitly investigated and quantified, either theoretically or via simulation, strengthening the quality of the end data product by taking out the guesswork. It is particularly convenient that the mathematical formulation of differential privacy employs the concept of a privacy loss budget, which acts as a fine-grained tuning parameter for the performance of the procedure. The framework is suited for integration with the *total budget* concept and the error decomposition approach to understanding the effect of individual error constituents. The price we pay for privacy can be regarded as a trade-off with the total utility, defined through concrete quality metrics on the resulting data product—for example, the minimal mean squared error achievable by an optimal survey design, or the accuracy on the output of certain routine data analysis protocols.

An increase in the PLB will in general improve the quality of the data product. But the impact on data quality exerted by a particular choice of PLB should be understood within the specific context of application. When the important use cases and accuracy targets are identified, transparency allows for the setting of privacy parameters to meet these targets via theoretical or simulated explorations, as early as during the design phase of the survey. As an illustration, Figure 2 repeats the same regression analysis as in Figure 1, but with $\epsilon = 1$, a PLB that is four times larger. While the correct, MCEM-based analysis remains valid, the naïve analysis has greatly improved its performance, as seen from the confidence ellipses in the left panel with comparable coverage compared to the right panel (correct analysis with $\epsilon = 1$), which is better than the middle panel of Figure 1 (naïve analysis with $\epsilon = 0.25$). Through six iterations of the 2010 Demonstration Data Files, the Census Bureau increased the PLB from $\epsilon = 6$, with $4$ for persons and $2$ for housing units (U.S. Census Bureau, 2019), to an equivalent of $(\epsilon, \delta) = (19.71, 10^{-10})$ for the production setting of the P.L. 94-171 files (U.S. Census Bureau, 2021c).[7] Since the PLB is a probabilistic bound on the log scale, a more than three-fold increase substantially weakened the privacy guarantee, but it allowed the bureau to improve and meet the various accuracy targets identified by the data user communities (U.S. Census Bureau, 2021b).

When privatization is a transparent procedure, it does not merely add to the total error of an otherwise confidential survey. We have reasons to hope that it may help reduce the error via means of human psychology. A primary cause of inaccuracy in the census is nonresponse and imperfect coverage, in part having to do with insufficient public trust, both in the privacy protection of disseminated data products and in the Census Bureau's ability to maintain confidentiality of sensitive information (boyd & Sarathy, 2022; Singer et al., 1993; Sullivan, 2020). Individual data contributors value their privacy. Through their data sharing (or rather, un-sharing) decisions, they exhibit a clear preference for privacy, which has both been theoretically studied (Ghosh & Roth, 2015; Nissim et al., 2012)and empirically measured (Acquisti et al., 2013). To the privacy-conscious data contributor, transparent privacy offers the certainty of knowing that our information is protected in an explicit and provable way that is vetted by communities of interested data users. In addition, transparent privacy enables a quantitative description of how the information from each data contributor supports fair and accurate policy decisions, which directly affect the welfare of individual respondents. Even a small progress toward instilling confidence and encouraging participation can reduce the potentially immense cost due to systematic nonresponse bias, and enhance the quality of the survey (Meng, 2018).

The algorithmic construction of differential privacy and the theoretical explorations of total survey error creates a promising intersection. We hope to see synergistic methodological developments to serve the dual purpose of efficient privacy protection and survey quality optimization. I will briefly discuss one such direction. Discussing *TSE-aware SDL*, Karr (2017) advocates that when additive privacy mechanisms are employed, the optimal choice of privacy error covariance should accord to the measurement error covariance. The resulting data release demonstrates superior utility in terms of closeness to the confidential data distribution in the sense of minimal Kullbeck-Leibler divergence. This proposal, when accepted into the differential privacy framework, requires generalizing the vanilla algorithms to produce correlated noise while preserving the privacy guarantee. Differential privacy researchers have looked in this direction and offered tools adaptable to this purpose. For example, Nikolov et al. (2013) propose a correlated Gaussian mechanism for linear queries, and demonstrate that it is an optimal mechanism among $(\epsilon, \delta)$-differentially private mechanisms in terms of minimizing the mean squared error of the data product. A privacy mechanism structurally designed to express the theory of survey error minimization paves the way for optimized usability of the end data product.

## 6. The Quest for Full Transparency: Are We There Yet?

The collection of economic and social data is a widely practiced tradition in many civilizations, which traces back hundreds if not thousands of years. It was not until the latter part of the 20th century, however, that the need to defend individuals' confidentiality became recognized as a worthy scholarly pursuit (Oberski & Kreuter, 2020). Despite privacy being a youthful subject, we have come a long way in a mere couple of decades to advance the art and the science of privacy protection. The progress was driven by a series of embarrassments (some mentioned in Section 1), an awareness shared by major data curators including official statistics agencies, corporations, and research institutions, and most importantly the hard work of computer scientists and statisticians who keep inventing new techniques to replace the old. Transparent privacy is a significant milestone in this progress, a gift bestowed upon us by the continued advancement in privacy research. However, a perpetually curious researcher still must ask the ungrateful question: is this transparency all we can ask for?

Just as some gifts are more practical than others, some versions of transparent privacy are more usable than others. An example of transparent privacy that can be difficult to work with occurs when constraints—including *invariants*, nonnegativity, integer characteristics, and structural consistencies— must be simultaneously imposed on the differentially private queries.

Invariants are a set of exact statistics calculated based on the confidential microdata (Abowd et al., 2022; Ashmead et al., 2019). Some invariants are mandated, in the sense that all versions of the privatized data that the curator can release must accord to these values. Invariants represent use cases for which a precise enumeration is crucial. For example, the total population of each state, which serves as the basis for the allocation of House seats, must be reported exactly as enumerated as required by the U.S. Constitution.

What information is deemed invariant, and what characteristics of the confidential data should form constraints on the privatized data are ultimately a policy decision. However, constraints don't mingle with classical differential privacy in a straightforward manner. Indeed, if a query has unbiased random noise added to it, there is no guarantee that it still possesses the same characteristics as does the noiseless version. The task of ensuring privatized census data releases to be constraint-complaint is performed by the TopDown Algorithm (Abowd et al., 2022). The algorithm consists of two phases. During the *measurement* phase, differentially private *noisy measurements*, which are counts infused with unbiased discrete Gaussian noises, are generated for each geographic level. During the *estimation* phase, the algorithm employs nonnegative $L_2$ optimization followed by $L_1$ controlled rounding, to ensure that the output consists of only nonnegative integers while satisfying all desired constraints. It has been recognized that optimization-based postprocessing can create unexpected anomalies in the released tabulations, namely systematic positive biases for smaller counts and negative biases for larger counts, at a magnitude that tends to overwhelm the amount of inaccuracy due to privacy alone (Devine et al., 2020; Zhu et al., 2021).

Due to the sheer size of the optimization problem, the statistical properties of its output do not succumb easily to theoretical explorations. However, the observed adverse effects of such processing should not strike us as unanticipated. Projective optimizations, be they $L_2$ or $L_1$, are essentially regression adjustments on a collection of data points. The departures that the resulting values exhibit in the direction opposite to the original values is a manifestation of the Galtonian phenomenon of *regression toward the mean* (Stigler, 2016). Furthermore, whenever an unbiased and unbounded estimator is a posteriori confined to a subdomain (the nonnegative integers), the unbiasedness property it once enjoys may no longer hold (Berger, 1990).

Note that an optimization algorithm that imposes invariants can still be *procedurally* transparent. The design of the TopDown Algorithm is documented in the Census Bureau's publication (Abowd et al., 2022), accompanied by a suite of demonstration products and the GitHub codebase (2020 Census DAS Development Team, 2021). However, mere procedural transparency may not be good enough. In summary of the NASEM CNSTAT workshop dedicated to the assessment of the 2020 Census DAS, Hotz and Salvo (2022) note that postprocessing of privatized data can be particularly difficult to model statistically. This is because the optimization imposes an extremely complex, indeed data-dependent, function to the confidential data (Gong & Meng, 2020). As a result, the distributional description of the overall algorithm (including postprocessing), denoted as $p_\xi$ in this article, is difficult to characterize. One might still be able to draw limited inferential conclusions by invoking certain approximate or robustness arguments (see e.g. Avella-Medina, 2021;

Dimitrakakis et al., 2014; Dwork & Lei, 2009). However, if the statistical properties of the end data release cannot be simply described or replicated on an ordinary personal computer, it sets back the transparency brought forth by the differentially private noise-infusion mechanism, and hinders a typical end user's ability to carry out the principled analysis according to (4.2), (4.3), or (4.5), as Section 4 outlines.

Nevertheless, procedural transparency is a promising step toward the full transparency that is needed to support principled statistical inference. Through the design phase of the 2020 DAS for the P.L. 94-171 data products, the Census Bureau released a total of six rounds of demonstration data files in the form of privacy-protect microdata files (PPMFs). The PPMFs enabled community assessments on the DAS performance, including its accuracy targets, and to provide feedback to the Census Bureau for future improvement. These demonstration data are a crucial source of information for the data-user communities, and have supported research on the impact of differential privacy as well as postprocessing in topics such as small area population (Swanson et al., 2021; Swanson & Cossman, 2021), tribal nations (National Congress of American Indians, 2021), redistricting and voting rights measures (Cohen et al., 2022; Kenny et al., 2021).

On August 12, 2021, a group of privacy researchers signed a letter addressed to Dr. Ron Jarmin, Acting Director of the United States Census Bureau, to request the release of the noisy measurement files that accompanied the P.L. 94-171 redistricting data products (Dwork et al., 2021). The letter made the compelling case that the noisy measurement files present the most straightforward solution to the issues that arise due to postprocessing. Since the noisy measurements are already formally private, releasing these files does not pose an additional threat to the privacy guarantee that the Bureau already offers. On the other hand, they will allow researchers to quantify the biases induced by postprocessing and to conduct correct uncertainty quantification. In the report *Consistency of Data Products and Formal Privacy Methods for the 2020 Census*, JASON (2022, p.8) makes the recommendation that the Bureau "should not reduce the information value of their data products solely because of fears that some stakeholders will be confused by or misuse the released data." It makes an explicit call for the release of all noisy measurements used to produce the released data products that do not unduly increase disclosure risk, and the quantification of uncertainty associated with the publicized data products. On April 28–29, 2022, a workshop dedicated to articulating a technical research agenda for statistical inference on the differentially private census noisy measurement files took place at Rutgers University, gathering experts from domains of social sciences, demography, public policy, statistics, and computer science. These efforts reflect the shared recognition among the research and policy communities that access to the census noisy measurement files, and its associated transparency benefits, are both crucial and feasible within the current disclosure avoidance framework that the Census Bureau employs.

The evolution of privacy science over the years reflects the growing dynamic among several branches of data science, as they collectively benefit from vastly improved computational and data storage abilities. What we're witnessing today is a paradigm shift in the science of curating official, social, and personal statistics. A change of this scale is bound to exert seismic impact on the ways that quantitative evidence is used and interpreted, raising novel questions and opportunities in all disciplines that rely on these data sources. The protection of privacy is not just a legal or policy mandate, but an ethical treatment of all individuals who contribute to the collective betterment of science and society with their information. As privacy research continues to evolve, an open and cross-disciplinary conversation is the catalyst to a fitting solution. Partaking in this conversation is our opportunity to defend democracy in its modern form: underpinned by numbers, yet elevated by our respect for one another as more than just numbers.

---

## Acknowledgments

Ruobin Gong wishes to thank Xiao-Li Meng for helpful discussions, and five anonymous reviewers for their comments.

## Disclosure Statement

Ruobin Gong's research is supported in part by the National Science Foundation (DMS-1916002).

---

---

# Appendices

## Appendix A: Analytical Form of the Biasing Effect in Large Finite Samples

Here we state a central limit theorem for the naïve slope estimator $\hat{b}_1$, applicable when the independent variable $x_i$'s are treated as fixed and when the sample size is large.

**Theorem 2**. ***Let*** $v_n^x = \frac{1}{n}\sum_{i=1}^{n}(x_i - \bar{x})^2$ ***and*** $k_n^x = \frac{1}{n}\sum_{i=1}^{n}(x_i - \bar{x})^4$ ***respectively denote the (unadjusted) sample variance and kurtosis of the confidential data*** $\{x_i\}_{i=1}^n$***. Assume*** $\lim_{n\to\infty} k_n^x = k > 0$ ***is well-defined. Privatized data*** $(\tilde{x}_i, \tilde{y}_i)$ ***follows the generative model in (3.1) and privacy mechanism in (3.2). The naïve slope estimator for the simple linear regression of*** $\tilde{y}_i$ ***against*** $\tilde{x}_i$ ***is*** $\hat{b}_1 = \sum_{i=1}^{n}(\tilde{x}_i - \bar{\tilde{x}})(\tilde{y}_i - \bar{\tilde{y}})/\sum_{i=1}^{n}(\tilde{x}_i - \bar{\tilde{x}})^2$***. Then, as*** $n \to \infty$***,***

$$\sqrt{n}\left(\frac{\hat{b}_1 - \gamma_n\beta_1}{\sqrt{\tilde{\sigma}_n}}\right) \xrightarrow{d} N(0,1), \tag{A.1}$$

***where*** $\gamma_n = v_n^x / \left(v_n^x + \sigma_u^2\right)$ ***is the biasing coefficient, and***

$$\tilde{\sigma}_n = \frac{\beta_1^2\left[\gamma_n^2\left(k_n^x + 6\sigma_u^2 v_n^x + 6\sigma_u^4\right) - 2\gamma_n\left(k_n^x + 3\sigma_u^2 v_n^x\right) + k_n^x + \sigma_u^2 v_n^x\right] + \left(\sigma_v^2 + \sigma^2\right)\left(v_n^x + \sigma_u^2\right)}{\left(v_n^x + \sigma_u^2\right)^2}$$

***the approximate standard error.***

The biasing coefficient $\gamma_n$ is the finite-population counterpart to the ratio $\mathbb{V}(x) / \left(\mathbb{V}(x) + \sigma_u^2\right)$ discussed in Section 3. As a special case when no privacy protection is performed on either $x_i$ or $y_i$, that is, $\sigma_u^2 = \sigma_v^2 = 0$, then the biasing coefficient $\gamma_n = 1$, and the associated variance $\tilde{\sigma}_n = \sigma^2 / v_n^x$ regardless of sample size $n$. This recovers the usual sampling result for the classic regression estimate $\hat{\beta}_1$. Otherwise when $\sigma_u^2 > 0$, the biasing coefficient $\gamma_n$ is a positive fractional quantity, tending towards $0$ as $\epsilon_x$ decreases, and $1$ if it increases. Therefore, the naïve estimator $\hat{b}_1$ underestimates the strength of association between $x$ and $y$, more severely so as the privacy protection for $x$ becomes more stringent.

When $n$ is large, the large sample sampling distribution of $\hat{b}_1$ has $(1-\alpha)\%$ of its mass within the lower and upper distribution limits $\left(\gamma_n\beta_1 - \Phi(1-\alpha/2)\sqrt{\tilde{\sigma}_n/n}, \gamma_n\beta_1 + \Phi(1-\alpha/2)\sqrt{\tilde{\sigma}_n/n}\right)$, which are functions of the true $\beta_1$, the confidential data $\{x_i\}_{i=1}^n$, as well as the idiosyncratic variance ($\sigma^2$) and the privacy error variances ($\sigma_u^2$ and $\sigma_v^2$). The left panel of Figure A.1 depicts these large sample $95\%$ distribution limits under various privacy-loss budget settings for $x$ and $y$, and the right panel depicts their actual coverage probability for the true parameter $\beta_1$.

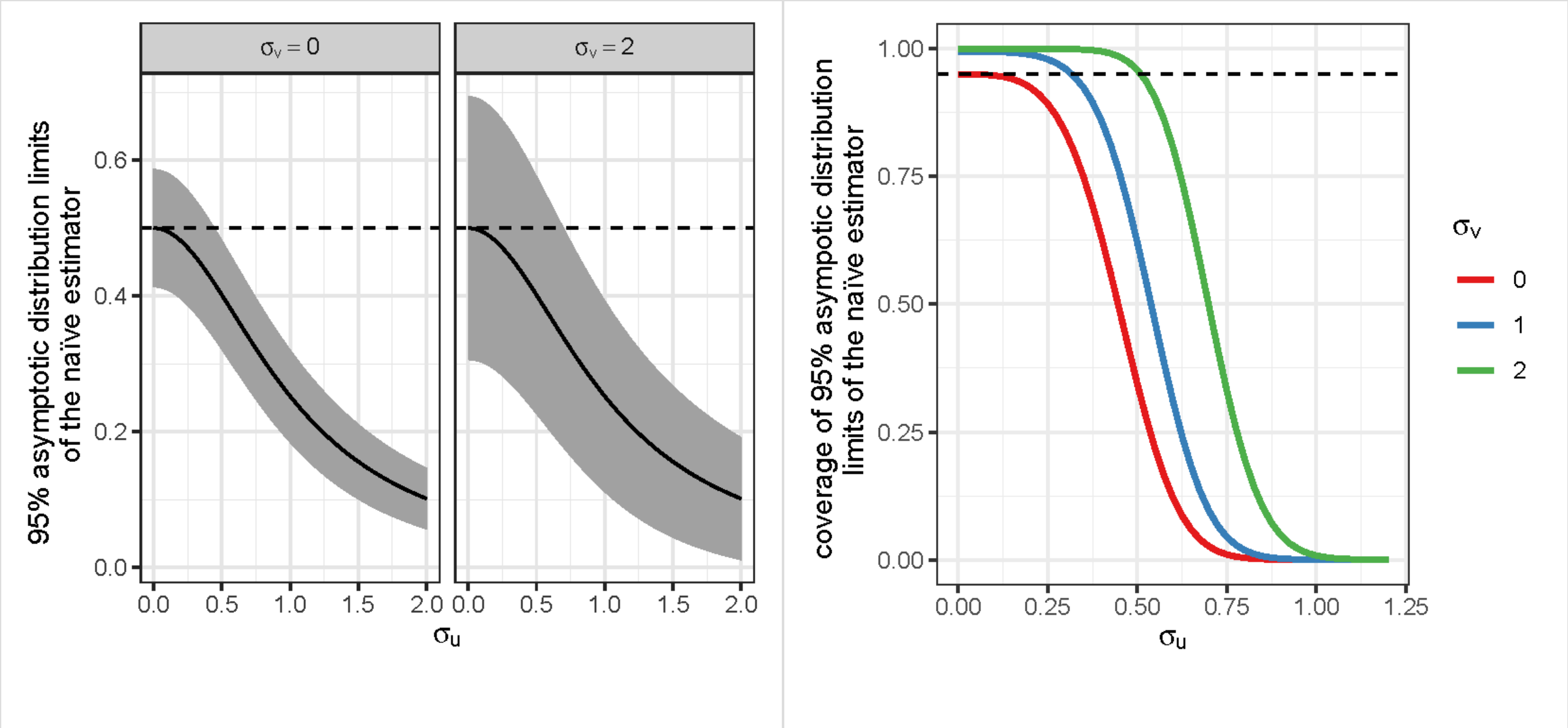

**Figure A.1.** Biasing Effect of privacy noise in linear regression. Left: large sample 95% distribution limits of the naïve slope estimator $\hat{b}_1$ as a function of $\sigma_u$ and $\sigma_v$ (privacy error variances of $x$ and $y$, respectively). The panel labeled "$\sigma_v = 0$" shows distribution limits (shaded gray) around the point-wise limit of the naïve estimator (black solid line), if $y$ is not privacy protected but $x$ is protected at increasing levels of stringency (as much as $\sigma_u = \sqrt{2}/\epsilon_x = 2$ or $\sigma_x = 0.707$). The panel labeled "$\sigma_v = 2$" shows distribution limits if $y$ is also protected at that scale (equivalent to $\epsilon_y = 0.707$). True $\beta_1 = 0.5$ (black dashed line). Right: coverage probabilities of the large sample 95% distribution limits for the naïve slope estimator $\hat{b}_1$, as a function of $\sigma_u$ and $\sigma_v$. With no privacy protection for either $x$ or $y$ $(\sigma_u = \sigma_v = 0)$, the 95% distribution limit coincides with that of $\hat{\beta}_1$ from the classic regression setting, and meets its nominal coverage for all $n$. Adding privacy protection to $y$ only (i.e., $\sigma_v$ increases) inflates a correctly centered asymptotic distribution, exhibiting conservative coverage. However for fixed $\sigma_v$, as privacy protection for $x$ increases (i.e., $\sigma_u$ increases), the bias in $\hat{b}_1$ dominates and drives coverage probability down to zero. Illustration uses a data set of $n = 500$, with sample variance of confidential $x$ about 1.023, and idiosyncratic error variance $\sigma^2 = 1$.

We now supply the proof of Theorem 2, which gives a large sample approximation to the distribution of the naiïve regression slope estimator for privatized data, which takes the form of

$$\hat{b}_1 = \frac{\sum_{i=1}^n (\tilde{x}_i - \bar{\tilde{x}})(\tilde{y}_i - \bar{\tilde{y}})}{\sum_{i=1}^n (\tilde{x}_i - \bar{\tilde{x}})^2}$$
$$= \frac{\sum_{i=1}^n ((x_i - \bar{x}) + (u_i - \bar{u}))(\beta_1 (x_i - \bar{x}) + (v_i - \bar{v}) + (e_i - \bar{e}))}{\sum_{i=1}^n ((x_i - \bar{x}) + (u_i - \bar{u}))^2}$$

Writing $c_i = v_i + e_i$ and $a_i = x_i - \bar{x}$, we have that

$$\hat{b}_1 = \frac{\sum_{i=1}^n (a_i + u_i - \bar{u})(\beta_1 a_i + c_i - \bar{c})}{\sum_{i=1}^n (a_i + u_i - \bar{u})^2} = \frac{\frac{1}{n}\sum_{i=1}^n (a_i + u_i)(\beta_1 a_i + c_i - \bar{c})}{\frac{1}{n}\sum_{i=1}^n (a_i + u_i)^2 - \bar{u}^2} =: \frac{A_n}{B_n}$$

Using independence between $c_i$ and $u_i$, denoting the sample variance and kurtosis of $x$ as

$$v_n^x := \frac{1}{n}\sum_{i=1}^n a_i^2, \qquad k_n^x := \frac{1}{n}\sum_{i=1}^n a_i^4,$$

assuming that $\lim_{n\to\infty} k_n^x = k > 0$ exists and is well-defined. We have that by law of large numbers,

$$A_n - \beta_1 v_n^x \xrightarrow{p} 0, \qquad B_n - \left(v_n^x + \sigma_u^2\right) \xrightarrow{p} 0,$$

thus

$$\hat{b}_1 \xrightarrow{p} \gamma_n \beta_1$$

where $\gamma_n = \frac{v_n^x}{v_n^x + \sigma_u^2}$ is the biasing coefficient for the naive slope estimator $\hat{b}_1$. To establish the Central Limit Theorem result, let us first consider

$$\begin{aligned} A_n' = & \quad \frac{1}{n}\sum_{i=1}^n (a_i + u_i)(\beta_1 a_i + c_i) = A_n - \bar{c}\bar{u}, \\ B_n' = & \quad \frac{1}{n}\sum_{i=1}^n (a_i + u_i)^2 = B_n + \bar{u}^2. \end{aligned}$$

We have that

$$\sqrt{n}\left(A_n - \gamma_n\beta_1 B_n\right) = \sqrt{n}\left(A_n' - \gamma_n\beta_1 B_n'\right) + \sqrt{n}\bar{c}\bar{u} - \sqrt{n}\gamma_n\beta_1\bar{u}^2,$$

where $\sqrt{n}\bar{c}\bar{u} \xrightarrow{p} 0$ and $\sqrt{n}\gamma_n\beta_1\bar{u}^2 \xrightarrow{p} 0$.The following central limit theorem holds:

$$\begin{aligned} \sqrt{n}\left(\frac{A_n' - \gamma_n\beta_1 B_n'}{\sqrt{\Sigma_n}}\right) = & \quad \frac{1}{\sqrt{n\Sigma_n}}\left[\sum_{i=1}^n (a_i + u_i)(\beta_1 a_i + c_i) - \gamma_n\beta_1 \sum_{i=1}^n (a_i + u_i)^2\right] \\ \xrightarrow{d} & \quad N(0,1) \end{aligned}$$

where

$$\begin{aligned} \Sigma_n = & \quad \frac{1}{n}\sum_{i=1}^n E\left((a_i + u_i)(\beta_1 a_i + c_i) - \gamma_n\beta_1 (a_i + u_i)^2\right)^2 \\ = & \quad \gamma_n^2\beta_1^2\left(\frac{1}{n}\sum a_i^4 + 6\sigma_u^2\frac{1}{n}\sum a_i^2 + 6\sigma_u^4\right) - 2\gamma_n\beta_1^2\left(\frac{1}{n}\sum a_i^4 + 3\sigma_u^2\frac{1}{n}\sum a_i^2\right) + \\ & \quad \beta_1^2\left(\frac{1}{n}\sum a_i^4 + \sigma_u^2\frac{1}{n}\sum a_i^2\right) + \left(\sigma_v^2 + \sigma^2\right)\left(\frac{1}{n}\sum a_i^2 + \sigma_u^2\right) \\ = & \quad \beta_1^2\left[\gamma_n^2\left(k_n^x + 6\sigma_u^2 v_n^x + 6\sigma_u^4\right) - 2\gamma_n\left(k_n^x + 3\sigma_u^2 v_n^x\right) + k_n^x + \sigma_u^2 v_n^x\right] + \left(\sigma_v^2 + \sigma^2\right)\left(v_n^x + \sigma_u^2\right), \end{aligned}$$

noting that for each $i$,

$$\begin{aligned} & E\left(\left(a_i+u_i\right)\left(\beta_1 a_i+c_i\right)-\gamma_n \beta_1\left(a_i+u_i\right)^2\right)^2 \\ = \quad & E\left(\left(a_i+u_i\right)\left(\beta_1 a_i+c_i\right)\right)^2+\gamma_n^2 \beta_1^2 E\left(a_i+u_i\right)^4-2 \gamma_n \beta_1 E\left(\beta_1 a_i+c_i\right)\left(a_i+u_i\right)^3, \\ = \quad & E\left(\beta_1 a_i+c_i\right)^2 E\left(a_i+u_i\right)^2+\gamma_n^2 \beta_1^2 E\left(a_i+u_i\right)^4-2 \gamma_n \beta_1^2 a_i E\left(a_i+u_i\right)^3 \\ = \quad & \gamma_n^2 \beta_1^2\left(a_i^4+6 a_i^2 \sigma_u^2+6 \sigma_u^4\right)-2 \gamma_n \beta_1^2\left(a_i^4+3 a_i^2 \sigma_u^2\right)+\left(\beta_1^2 a_i^2+\sigma_v^2+\sigma^2\right)\left(a_i^2+\sigma_u^2\right), \end{aligned}$$

where for $u_i$ a centralized Laplace variable,

$$E\left(a_i+u_i\right)^2=a_i^2+\sigma_u^2; \qquad E\left(a_i+u_i\right)^3=a_i^3+3 a_i \sigma_u^2; \qquad E\left(a_i+u_i\right)^4=a_i^4+6 a_i^2 \sigma_u^2+6 \sigma_u^4.$$

Thus with $\hat{b}_1=A_n / B_n$, we have that the Central Limit Theorem for the naive slope estimator is

$$\sqrt{n}\left(\frac{\hat{b}_1-\gamma_n \beta_1}{\sqrt{\tilde{\sigma}_n}}\right)=\frac{\sqrt{n}\left(A_n^{\prime}-\gamma_n \beta_1 B_n^{\prime}\right)+\sqrt{n} \bar{c} \bar{u}-\sqrt{n} \gamma_n \beta_1 \bar{u}^2}{B_n \sqrt{\tilde{\sigma}_n}} \xrightarrow{d} N(0,1)$$

where

$$\begin{aligned} \tilde{\sigma}_n= & \left(v_n^x+\sigma_u^2\right)^{-2} \Sigma \\ = & \frac{\beta_1^2\left[\gamma_n^2\left(k_n^x+6 \sigma_u^2 v_n^x+6 \sigma_u^4\right)-2 \gamma_n\left(k_n^x+3 \sigma_u^2 v_n^x\right)+k_n^x+\sigma_u^2 v_n^x\right]+\left(\sigma_v^2+\sigma^2\right)\left(v_n^x+\sigma_u^2\right)}{\left(v_n^x+\sigma_u^2\right)^2}. \end{aligned}$$

As a special case when no privacy protection is performed on either $x$ or $y$, i.e. $\sigma_u=\sigma_v=0$, then $\gamma_n=1$ for all $n$ and $\tilde{\sigma}_n=\sigma^2 / v_n^x$ gives the usual sampling distribution result for $\hat{\beta}_1$.

## Appendix B: Equivalence Between (4.3) and (4.5)

The true posterior distribution in (4.3) is fully spelled out as

$$\begin{aligned} \pi_{\xi}(\beta \mid \tilde{s})= & \frac{\int \pi(\beta, s, \tilde{s}) d s}{\iint \pi(\beta, s, \tilde{s}) d s d \beta} \\ = & \frac{\pi_0(\beta) \int p_{\xi}(\tilde{s} \mid s) \mathcal{L}(\beta ; s) d s}{\int \pi_0(\beta) \int p_{\xi}(\tilde{s} \mid s) \mathcal{L}(\beta ; s) d s d \beta} \\ = & \frac{\pi_0(\beta) \mathcal{L}_{\xi}(\beta ; \tilde{s})}{\int \pi_0(\beta) \mathcal{L}_{\xi}(\beta ; \tilde{s}) d \beta}. \end{aligned}$$

Noting that

$$\pi(\beta \mid s)=\frac{\pi_0(\beta) \mathcal{L}(\beta ; s)}{\int \pi_0(\beta) \mathcal{L}(\beta ; s) d \beta}$$

and

$$\begin{aligned}\pi_\xi\left(s \mid \tilde{s}\right) = & \frac{\int \pi_0\left(\beta\right) p_\xi\left(\tilde{s} \mid s\right) \mathcal{L}\left(\beta; s\right) d\beta}{\iint \pi_0\left(\beta\right) p_\xi\left(\tilde{s} \mid s\right) \mathcal{L}\left(\beta; s\right) d\beta ds} \\ = & \frac{p_\xi\left(\tilde{s} \mid s\right) \int \pi_0\left(\beta\right) \mathcal{L}\left(\beta; s\right) d\beta}{\int p_\xi\left(\tilde{s} \mid s\right) \int \pi_0\left(\beta\right) \mathcal{L}\left(\beta; s\right) d\beta ds},\end{aligned}$$

we have that the right hand side of (4.5)

$$\begin{aligned}\int \pi\left(\beta \mid s\right) \pi_\xi\left(s \mid \tilde{s}\right) ds = & \int \frac{\pi_0\left(\beta\right) \mathcal{L}\left(\beta; s\right)}{\int \pi_0\left(\beta\right) \mathcal{L}\left(\beta; s\right) d\beta} \cdot \frac{p_\xi\left(\tilde{s} \mid s\right) \int \pi_0\left(\beta\right) \mathcal{L}\left(\beta; s\right) d\beta}{\int p_\xi\left(\tilde{s} \mid s\right) \int \pi_0\left(\beta\right) \mathcal{L}\left(\beta; s\right) d\beta ds} ds \\ = & \frac{\int p_\xi\left(\tilde{s} \mid s\right) \pi_0\left(\beta\right) \mathcal{L}\left(\beta; s\right) ds}{\int p_\xi\left(\tilde{s} \mid s\right) \int \pi_0\left(\beta\right) \mathcal{L}\left(\beta; s\right) d\beta ds} \\ = & \frac{\pi_0\left(\beta\right) \int p_\xi\left(\tilde{s} \mid s\right) \mathcal{L}\left(\beta; s\right) ds}{\int \pi_0\left(\beta\right) \int p_\xi\left(\tilde{s} \mid s\right) \mathcal{L}\left(\beta; s\right) ds d\beta} \\ = & \frac{\pi_0\left(\beta\right) \mathcal{L}_\xi\left(\beta; \tilde{s}\right)}{\int \pi_0\left(\beta\right) \mathcal{L}_\xi\left(\beta; \tilde{s}\right) d\beta} = \pi_\xi\left(\beta \mid \tilde{s}\right),\end{aligned}$$

establishing the equivalence between (4.3) and (4.5).

## Appendix C: Details of the MCEM Algorithm

The Monte Carlo expectation maximization (MCEM) via importance sampling algorithm works as follows for the linear regression example. The data generative mechanism is

$$x_i \sim Pois\left(10\right) \text{ i.i.d.}, \qquad y_i = -5 + 4x_i + e_i,\ e_i \sim N\left(0, \sigma^2 = 5^2\right),$$

followed by additive privatization

$$\tilde{x}_i = x_i + u_i, \qquad \tilde{y}_i = y_i + v_i,\ u_i, v_i \sim Lap\left(\epsilon^{-1}\right).$$

The goal is to estimate the parameter values, here set at $\beta_0 = -5, \beta_1 = 4$, with maximum likelihood estimation.

At the E-step, approximate the conditional expectation of the complete data likelihood with respect to the observed data and the parameter estimate at the current iteration

$$Q(\boldsymbol{\beta}; \boldsymbol{\beta}^{(t)}) = \mathbb{E}\left(\log \mathcal{L}\left(\boldsymbol{\beta}; s, \tilde{s}\right) \mid \tilde{s}, \boldsymbol{\beta}^{(t)}\right) = \mathbb{E}\left(\log \mathcal{L}(\boldsymbol{\beta}; s) \mid \tilde{s}, \boldsymbol{\beta}^{(t)}\right) + \text{const.}$$

where the log likelihood of the missing data is

$$\log \mathcal{L}(\boldsymbol{\beta}; s) = -\tfrac{1}{2\sigma^2} \sum_i \left(y_i - \beta_0 - \beta_1 x_i\right)^2,$$

with score

$$\mathsf{S}\left(\boldsymbol{\beta}; s\right) = \tfrac{\partial}{\partial \beta} \log \mathcal{L}\left(\boldsymbol{\beta}; s\right) = \tfrac{1}{\sigma^2} \begin{pmatrix} \sum_i y_i - n\beta_0 - \beta_1 \sum_i x_i \\ \sum_i x_i y_i - \beta_0 \sum_i x_i - \beta_1 \sum_i x_i^2 \end{pmatrix}$$

and negative second derivative

$$i(\boldsymbol{\beta}; s) = -\frac{\partial^2}{\partial \boldsymbol{\beta}^2} \log \mathcal{L}(\boldsymbol{\beta}; s) = \frac{1}{\sigma^2} \begin{pmatrix} n & \sum_i x_i \\ \sum_i x_i & \sum_i x_i^2 \end{pmatrix}.$$

The approximation to the $Q$ function is constructed with $k = 1, \ldots, K$ weighted samples of the confidential data likelihood:

$$\hat{Q}(\boldsymbol{\beta}; \boldsymbol{\beta}^{(t)}) = \frac{1}{\sum_k \omega_k} \sum_k \omega_k \log \mathcal{L}\left(\boldsymbol{\beta}; s_k^{(t)}\right),$$

where $s_k^{(t)} \sim \mathcal{L}\left(\boldsymbol{\beta}^{(t)}; s\right)$ consists of $x_{ik}^{(t)} \sim Pois(10)$ and $y_{ik}^{(t)} = \beta_0^{(t)} + \beta_1^{(t)} x_{ik}^{(t)} + e_{ik}^{(t)}$, where $e_{ik}^{(t)} \sim N\left(0, \sigma^2 = 5^2\right)$ for $i = 1, \ldots, 10$. The weights are calculated as

$$\omega_k = \prod_i f\left(x_{ik}^{(t)} - \tilde{x}_i; \epsilon^{-1}\right) f\left(y_{ik}^{(t)} - \tilde{y}_i; \epsilon^{-1}\right),$$

where $f(\cdot; b)$ is the Laplace density with scale parameter $b$.

The M-step, maximizing $\hat{Q}(\boldsymbol{\beta}; \boldsymbol{\beta}^{(t)})$, occurs at $\boldsymbol{\beta}^{(t+1)}$ which is the solution to the approximating score function being zero,

$$\begin{aligned} \frac{\partial}{\partial \boldsymbol{\beta}} \hat{Q}(\boldsymbol{\beta}; \boldsymbol{\beta}^{(t)}) &= \frac{\sum_k \omega_k \mathsf{S}\left(\boldsymbol{\beta}; s_k^{(t)}\right)}{\sum_k \omega_k} \\ &= \frac{1}{\sigma^2 \sum_k \omega_k} \sum_k \omega_k \begin{pmatrix} n\bar{y}_k^{(t)} - n\beta_0 - \beta_1 n \bar{x}_k^{(t)} \\ n\overline{(xy)}_k^{(t)} - \beta_0 n \bar{x}_k^{(t)} - \beta_1 n \overline{(x^2)}_k^{(t)} \end{pmatrix} = 0, \end{aligned}$$

where $\bar{y}_k^{(t)} = n^{-1} \sum_i y_{ik}^{(t)}$, $\overline{(xy)}_k^{(t)} = n^{-1} \sum_i x_{ik}^{(t)} y_{ik}^{(t)}$, and so on. Writing the $\omega$-weighted averages as

$$\bar{y}_\omega^{(t)} = \frac{\sum_k \omega_k \bar{y}_k^{(t)}}{\sum_k \omega_k}, \; \bar{x}_\omega^{(t)} = \frac{\sum_k \omega_k \bar{x}_k^{(t)}}{\sum_k \omega_k}, \; \overline{(xy)}_\omega^{(t)} = \frac{\sum_k \omega_k \overline{(xy)}_k^{(t)}}{\sum_k \omega_k}, \; \overline{(x^2)}_\omega^{(t)} = \frac{\sum_k \omega_k \overline{(x^2)}_k^{(t)}}{\sum_k \omega_k},$$

we have that

$$\beta_1^{(t+1)} = \frac{\overline{(xy)}_\omega^{(t)} - \bar{x}_\omega^{(t)} \bar{y}_\omega^{(t)}}{\overline{(x^2)}_\omega^{(t)} - \left(\bar{x}_\omega^{(t)}\right)^2}, \qquad \beta_0^{(t+1)} = \bar{y}_\omega^{(t)} - \beta_1^{(t+1)} \bar{x}_\omega^{(t)}, \tag{C.1}$$

which may be calculated at iteration $t$ to supply the parameter values for the next iteration $t + 1$. Furthermore, the observed Fisher information can be approximated as

$$\frac{\sum \omega_k i\left(\boldsymbol{\beta}; s_k^{(t)}\right)}{\sum_k \omega_k} - \frac{\sum \omega_k \mathsf{S}\left(\boldsymbol{\beta}; s_k^{(t)}\right) \mathsf{S}^\top\left(\boldsymbol{\beta}; s_k^{(t)}\right)}{\sum_k \omega_k} + \left(\frac{\sum_k \omega_k \mathsf{S}\left(\boldsymbol{\beta}; s_k^{(t)}\right)}{\sum_k \omega_k}\right) \left(\frac{\sum_k \omega_k \mathsf{S}\left(\boldsymbol{\beta}; s_k^{(t)}\right)}{\sum_k \omega_k}\right)^\top, \qquad \text{(C.2}$$

where the first term is equal to

$$\frac{\sum \omega_k i\left(\beta; s_k^{(t)}\right)}{\sum_k \omega_k} = \frac{1}{\sigma^2}\begin{pmatrix} 1 & \bar{x}_\omega^{(t)} \\ \bar{x}_\omega^{(t)} & \overline{(x^2)}_\omega^{(t)} \end{pmatrix},$$

the second equal to ("..'' denotes mirrored hence omitted entry in a symmetric matrix)

$$-\frac{n^2}{\sigma^4 \sum_k \omega_k}\begin{pmatrix} \sum_k \omega_k \left(\bar{y}_k^{(t)} - \beta_0 - \beta_1 \bar{x}_k^{(t)}\right)^2 & \sum_k \omega_k \left(\bar{y}_k^{(t)} - \beta_0 - \beta_1 \bar{x}_k^{(t)}\right)\left(\overline{(xy)}_k^{(t)} - \beta_0 \bar{x}_k^{(t)} - \beta_1 \overline{(x^2)}_k^{(t)}\right) \\ .. & \sum_k \omega_k \left(\overline{(xy)}_k^{(t)} - \beta_0 \bar{x}_k^{(t)} - \beta_1 \overline{(x^2)}_k^{(t)}\right)^2 \end{pmatrix},$$

and the third simply the outer product of the observed score from before.

The right panels of Figures 1 and 2 both follow the recipe outlined above to draw maximum likelihood inference for the regression demostration, using $\epsilon = 0.25$ and $\epsilon = 1$ respectively as the privacy loss budget. The 95% confidence ellipses (green) are derived using a large-sample normal approximation to the likelihood at the maximum likelihood estimate (MLE), with covariance equal to the inverse observed Fisher information centered at the MLE, obtained respectively according to (C.1) and (C.2) with the values of the parameters at the algorithm's convergence.

---

# Footnotes

1. Zero-concentrated differential privacy controls not the ratio, but the maximal *divergence*, between the probability distributions of the random query on neighboring databases. It delivers a more efficient PLB composition property compared to the basic one stated in the previous paragraph. ↩
2. The analysis presented in this section uses the basic composition property introduced in Section 2. The noise scales correspond to a privacy loss budget of $\epsilon_x = \sqrt{2}/\sigma_u$ and $\epsilon_y = \sqrt{2}/\sigma_v$ per coordinate. The statistical properties discussed here hinge on the choice of the noise family and the scale parameters $\sigma_u^2$ and $\sigma_v^2$ only. Superior composition (hence lower total PLB) may be achieved with better privacy mechanism design, although that is inconsequential to the analysis presented here. ↩
3. For example the Herfindahl-Hirschman index, which is often used as a measure of population diversity or market competitiveness. ↩
4. Note that the DAS also performs *postprocessing* in addition to noise infusion. The postprocessing step may introduce additional sources of error to the assessed value of *d*. Unfortunately, the effect of postprocessing may be difficult to describe analytically. Section 6 returns to postprocessing and its impact. ↩

5. A statistical estimation procedure is said to be self-efficient with respect to a *selection mechanism* (i.e., the privacy mechanism), if it cannot achieve a smaller mean squared error via application to the *selected* (i.e., privatized) data, instead of or in addition to the complete (i.e., confidential) data. See Definition 5 of Meng (1994). ↩

6. In particular, under MCAR the missingness model may only depend on observed covariates. Under MAR it may also depend on observed outcome variables. ↩

7. When conducting privacy loss accounting under zCDP, the budget parameter $\rho$ translates into a range of $(\epsilon, \delta)$ values. The production setting of the P.L. 94-171 files uses $\rho = 2.56$ and 0.07 for persons and housing unit tables respectively, reported as the equivalent of $(17.14, 10^{-10})$ and $(2.47, 10^{-10})$ in $(\epsilon, \delta)$ terms. See U.S. Census Bureau, 2021c; Section 2 of Abowd et al., 2022; and Bun & Steinke, 2016, for more details. ↩